\documentclass[conference]{IEEEtran}

\makeatletter
\def\ps@headings{%
\def\@oddhead{\mbox{}\scriptsize\rightmark \hfil \thepage}%
\def\@evenhead{\scriptsize\thepage \hfil \leftmark\mbox{}}%
\def\@oddfoot{}%
\def\@evenfoot{}}
\makeatother 

\usepackage[noadjust]{cite}
\usepackage{graphicx}
\usepackage{url}
\usepackage{amsmath}
\usepackage{threeparttable}

\begin{document}

\title{Understanding the Characteristics of Internet Short Video
Sharing: YouTube as a Case Study}
\author{\IEEEauthorblockN{Xu Cheng}
\IEEEauthorblockA{School of Computing Science\\
Simon Fraser University\\
Burnaby, BC, Canada\\
Email: xuc@cs.sfu.ca} \and \IEEEauthorblockN{Cameron Dale}
\IEEEauthorblockA{School of Computing Science\\
Simon Fraser University\\
Burnaby, BC, Canada\\
Email: camerond@cs.sfu.ca} \and \IEEEauthorblockN{Jiangchuan Liu}
\IEEEauthorblockA{School of Computing Science\\
Simon Fraser University\\
Burnaby, BC, Canada\\
Email: jcliu@cs.sfu.ca}}

\maketitle

\begin{abstract}

Established in 2005, YouTube has become the most successful Internet
site providing a new generation of short video sharing service.
Today, YouTube alone comprises approximately 20\% of all HTTP
traffic, or nearly 10\% of all traffic on the Internet.
Understanding the features of YouTube and similar video sharing
sites is thus crucial to their sustainable development and to
network traffic engineering.

In this paper, using traces crawled in a 3-month period, we present
an in-depth and systematic measurement study on the characteristics
of YouTube videos. We find that YouTube videos have noticeably
different statistics compared to traditional streaming videos,
ranging from length and access pattern, to their active life span,
ratings, and comments. The series of datasets also allows us to
identify the growth trend of this fast evolving Internet site in
various aspects, which has seldom been explored before.

We also look closely at the social networking aspect of YouTube, as
this is a key driving force toward its success. In particular, we
find that the links to related videos generated by uploaders'
choices form a small-world network. This suggests that the videos
have strong correlations with each other, and creates opportunities
for developing novel caching or peer-to-peer distribution schemes to
efficiently deliver videos to end users.

\end{abstract}


\section{Introduction}\label{SecIntro}

The recent two years have witnessed an explosion of networked video
sharing as a new killer Internet application. The most successful
site, YouTube, now features over 40 million videos and enjoys 20
million visitors each month \cite{YouTubeSurvey}. The success of
similar sites like GoogleVideo, YahooVideo, MySpace, ClipShack, and
VSocial, and the recent expensive acquisition of YouTube by Google,
further confirm the mass market interest. Their great achievement
lies in the combination of the content-rich videos and, equally or
even more importantly, the establishment of a social network. These
sites have created a video village on the web, where anyone can be a
star, from lip-synching teenage girls to skateboarding dogs. With no
doubt, they are changing the content distribution landscape and even
the popular culture \cite{YouTubeChanging}.

Established in 2005, YouTube is one of the fastest-growing websites,
and has become the 4th most accessed site in the Internet. It has a
significant impact on the Internet traffic distribution, and itself
is suffering from severe scalability constraints. Understanding the
features of YouTube and similar video sharing sites is crucial
to network traffic engineering and to sustainable development of
this new generation of service.

In this paper, we present an in-depth and systematic measurement
study on the characteristics of YouTube videos. We have crawled the
YouTube site for a 3-month period in early 2007, and have obtained
27 datasets totaling 2,676,388 videos. This constitutes a
significant portion of the entire YouTube video repository, and
because most of these videos are accessible from the YouTube
homepage in less than 10 clicks, they are generally active and thus
representative for measuring the repository. Using this collection
of datasets, we find that YouTube videos have noticeably different
statistics from traditional streaming videos, in aspects from video
length and access pattern, to life span. There are also new features
that have not been examined by previous measurement studies, for
example, the ratings and comments. In addition, the series of
datasets also allows us to identify the growth trend of this fast
evolving Internet site in various aspects, which has seldom been
explored before.

We also look closely at the social networking aspect of YouTube, as
this is a key driving force toward the success of YouTube and
similar sites. In particular, we find that the links to related
videos generated by uploader's choices form a small-world network.
This suggests that the videos have strong correlations with each
other, and creates opportunities for developing novel caching or
peer-to-peer distribution schemes to efficiently deliver videos to
end users.

The rest of the paper is organized as follows. Section \ref{SecBack}
presents some background information and other related work. Section
\ref{SecMethodology} describes our method of gathering information
about YouTube videos, which is analyzed generally in Section
\ref{SecChar}, while the social networking aspects are analyzed
separately in Section \ref{SecSocial}. Section \ref{SecImpli}
discusses the implications of the results, and suggests ways that
the YouTube service could be improved. Finally, Section
\ref{Conclusion} concludes the paper.


\section{Background and Related Work}\label{SecBack}

\subsection{Internet Video Sharing}

Online videos existed long before YouTube entered the scene.
However, uploading videos, managing, sharing and watching them was
very cumbersome due to a lack of an easy-to-use integrated platform.
More importantly, the videos distributed by traditional media
servers and peer-to-peer file downloads like BitTorrent were
standalone units of content. Each single video was not connected to
other related video clips, for example other episodes of a show that
the user had just watched. Also, there was very little in the way of
content reviews or ratings.

The new generation of video sharing sites, YouTube and its
competitors, overcame these problems. They allow content suppliers
to upload video effortlessly, automatically converting from many
different formats, and to tag uploaded videos with keywords. Users
can easily share videos by mailing links to them, or embedding
them on web pages or in blogs. Users can also rate and comment on
videos, bringing new social aspects to the viewing of videos.
Consequently, popular videos can rise to the top in a very organic
fashion.

The social network existing in YouTube further enables communities
and groups. Videos are no longer independent from each other, and
neither are users. This has substantially contributed to the success
of YouTube and similar sites.

\subsection{Workload Measurement of Traditional Media Servers}

There has been a significant research effort into understanding the
workloads of traditional media servers, looking at, for example, the
video popularity and access locality \cite{tang03long, almeida01edu,
acharya00characterizing, yu06understanding}. The different aspects
of media and web objects, and those of live and stored video streams
have also been compared \cite{chesire01measurement,
veloso02hierarchical}. We have found that, while sharing similar
features, many of the video statistics of these traditional media
servers are quite different from YouTube; for example, the video
length distribution and life span. More importantly, these
traditional studies lack a social network among the videos.

The most similar work to ours is the very recent study by Huang et.
al. \cite{huang07peervod}. They analyzed a 9-month trace of
\emph{MSN Video}, Microsoft's VoD service, examining the user
behavior and popularity distribution of videos. This analysis led to
a \emph{peer-assisted VoD} design for reducing the server's
bandwidth costs. The difference to our work is that MSN Video is a
more traditional video service, with much fewer videos, most of
which are longer than all YouTube videos. MSN Video also has no
listings of related videos or user information, and thus no social
networking aspect.


\section{Methodology of Measurement}\label{SecMethodology}

In this paper, we focus on the access patterns and social networks
present in YouTube. To this end, we have crawled the YouTube site
for a 3-month period and obtained information on its videos through
a combination of the YouTube API and scrapes of YouTube video web
pages. The results offer a series of representative partial
snapshots of the YouTube video repository as well as its changing
trends.

\subsection{Video Format and Meta-data}\label{SubsecMetadata}


YouTube's video playback technology is based on Macromedia's Flash
Player and uses the Sorenson Spark H.263 video codec with pixel
dimensions of 320 by 240 and 25 frames per second. This technology
allows YouTube to display videos with quality comparable to more
established video playback technologies (such as Windows Media
Player, Realplayer or Apple's Quicktime Player). YouTube accepts
uploaded videos in WMV, AVI, MOV and MPEG formats, which are
converted into .FLV (Adobe Flash Video) format after uploading
\cite{YouTubeVideo}. It has been recognized that the use of a
uniform easily-playable format has been a key in the success of
YouTube.

There are many ways that YouTube's service differs from a
traditional media server. YouTube's FLV videos are not streamed to
the user, but are instead downloaded over a normal HTTP connection.
They are also not rate controlled to the playback rate of the video
but are sent at the maximum rate that the server and user can
accomplish, and there is no user interactivity from the server's
point of view (except for possibly stopping the download). In order
to fast forward the user must wait for that part of the video to
download, and pausing the playback does not pause the download.

YouTube randomly assigns each video a distinct 64-bit number, which
is represented in base 64 by an 11-digit ID composed of 0-9, a-z,
A-Z, -, and \_. Each video contains the following intuitive
meta-data: user who uploaded it, date when it was uploaded,
category, length, number of views, number of ratings, number of
comments, and a list of ``related videos''. The related videos are
links to other videos that have a similar title, description, or
tags, all of which are chosen by the uploader. A video can have
hundreds of related videos, but the webpage only shows at most 20 at
once, so we limit our scrape to these top 20 related videos. A
typical example of the meta-data is shown in Table
\ref{YouTubeMeta}.

\begin{table}
  \centering
  \begin{tabular}{|l|l|}
    \hline
    \textbf{ID} & 2AYAY2TLves \\
    \hline
    \textbf{Uploader} & GrimSanto \\
    \hline
    \textbf{Added Date} & May 19, 2007 \\
    \hline
    \textbf{Category} & Gadgets \& Games \\
    \hline
    \textbf{Video Length} & 268 seconds \\
    \hline
    \textbf{Number of Views} & 185,615 \\
    \hline
    \textbf{Number of Ratings} & 546 \\
    \hline
    \textbf{Number of Comments} & 588 \\
    \hline
    \textbf{Related Videos} & aUXoekeDIW8,\\
    & Sog2k6s7xVQ, \ldots\\
    \hline
  \end{tabular}
  \caption{Meta-data of a YouTube Video}\label{YouTubeMeta}
\end{table}

\subsection{YouTube Crawler}\label{SubsecCrawler}

We consider all the YouTube videos to form a directed graph, where
each video is a node in the graph. If video $b$ is in the related
video list (first 20 only) of video $a$, then there is a directed
edge from $a$ to $b$. Our crawler uses a breadth-first search to
find videos in the graph. We define the initial set of 0-depth video
IDs, which the crawler reads in to a queue at the beginning of the
crawl. When processing each video, it checks the list of related
videos and adds any new ones to the queue. The crawler is
single-threaded to avoid being suspected of a network attack.

Given a video ID, the crawler first extracts information from the
YouTube API, which contains all the meta-data except date added,
category, and related videos. The crawler then scrapes the video's
webpage to obtain the remaining information.

Our first crawl was on February 22nd, 2007, and started with the
initial set of videos from the list of ``Recently Featured'', ``Most
Viewed'', ``Top Rated'' and ``Most Discussed'', for ``Today'',
``This Week'', ``This Month'' and ''All Time'', which totalled 189
unique videos on that day. The crawl went to
more than four depths (the fifth was not completed), finding
approximately 750 thousand videos in about five days.

In the following weeks we ran the the crawler every two to three
days, each time defining the initial set of videos from the list of
``Most Viewed'', ``Top Rated'', and ``Most Discussed'', for
``Today'' and ``This Week'', which is about 200 to 300 videos. On
average, the crawl finds 80 thousand videos each time in less than
10 hours.

To study the growth trend of the video popularity, we also use the
crawler to update the statistics of some previously found videos.
For this crawl we only retrieve the number of views for relatively
new videos (uploaded after February 15th, 2007). This crawl is
performed once a week from March 5th to April 16th 2007, which
results in seven datasets.

We also separately crawled the file size and bit-rate information.
To get the file size, the crawler retrieves the response information
from the server when requesting to download the video file and
extracts the information on the size of the download. Some videos
also have the bit-rate embedded in the FLV video meta-data, which
the crawler extracts after downloading the beginning of the video
file.

Finally, we have also collected some information about YouTube
users. The crawler retrieves information on the number of uploaded
videos and friends of each user from the YouTube API, for a total of
more than 1 million users.


\section{Characteristics of YouTube Video}\label{SecChar}

From the first crawling on February 22nd, 2007, to the end of April,
2007, we have obtained 27 datasets totaling 2,676,388 videos. This
constitutes a significant portion of the entire YouTube video
repository (there are an estimated 42.5 million videos on YouTube
\cite{VideoNumber}). Also, because most of these videos can be
accessed from the YouTube homepage in less than 10 clicks, they are
generally active and thus representative for measuring
characteristics of the repository.

In the measurements, some characteristics are static and can be
measured once from the entire dataset: e.g. category, length, and
date added. Some characteristics are dynamic and can change from
dataset to dataset: e.g. number of views, ratings, and comments. We
consider this dynamic information to be static over a single crawl.
Later, the updated number of views information will be used to
measure the growth trend and life span of videos.

\subsection{Video Category}\label{SubsecCate}

\begin{table}
  \centering
  \begin{tabular}{|l|l|l|}
    \hline
    \textbf{Category} & \textbf{Count} & \textbf{Percentage}\\
    \hline
    Autos \& Vehicles & 66878 & 2.5\%\\
    Comedy & 323814 & 12.1\%\\
    Entertainment & 475821 & 17.8\%\\
    Film \& Animation & 225817 & 8.4\%\\
    Gadgets \& Games & 196026 & 7.3\%\\
    Howto \& DIY & 53291 & 2.0\%\\
    Music & 613754 & 22.9\%\\
    News \& Politics & 116153 & 4.3\%\\
    People \& Blogs & 199014 & 7.4\%\\
    Pets \& Animals & 50092 & 1.9\%\\
    Sports & 258375 & 9.7\%\\
    Travel \& Places & 58678 & 2.2\%\\
    Unavailable & 24068 & 0.9\%\\
    Removed & 14607 & 0.5\%\\
    \hline
  \end{tabular}
  \caption{List of YouTube video categories}\label{CategoryTable}
\end{table}

\begin{figure}
\centering
\includegraphics[width=0.8\columnwidth]{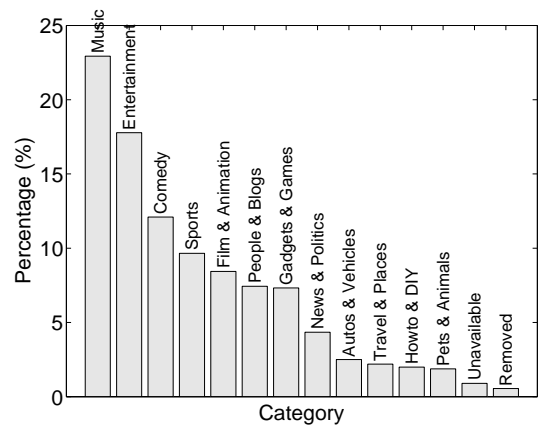}
\caption{Distribution of YouTube Videos' Categories}
\label{CategoryBar}
\end{figure}

One of 12 categories is selected by the user when uploading the
video. Table \ref{CategoryTable} lists the number and percentage of
all the categories, which is also shown graphically in Figure
\ref{CategoryBar}. In our entire dataset we can see that the
distribution is highly skewed: the most popular category is Music,
at about 22.9\%; the second is Entertainment, at about 17.8\%; and
the third is Comedy, at about 12.1\%.

In the table, we also list two other categories. ``Unavailable'' are
videos set to private, or videos that have been flagged as
inappropriate video, which the crawler can only get information for
from the YouTube API. ``Removed'' are videos that have been deleted
by the uploader, or by a YouTube moderator (due to the violation of
the terms of use), but still are linked to by other videos.

\begin{figure*}
\center
\begin{minipage}{5.5cm}
\centering
\includegraphics[width=\textwidth]{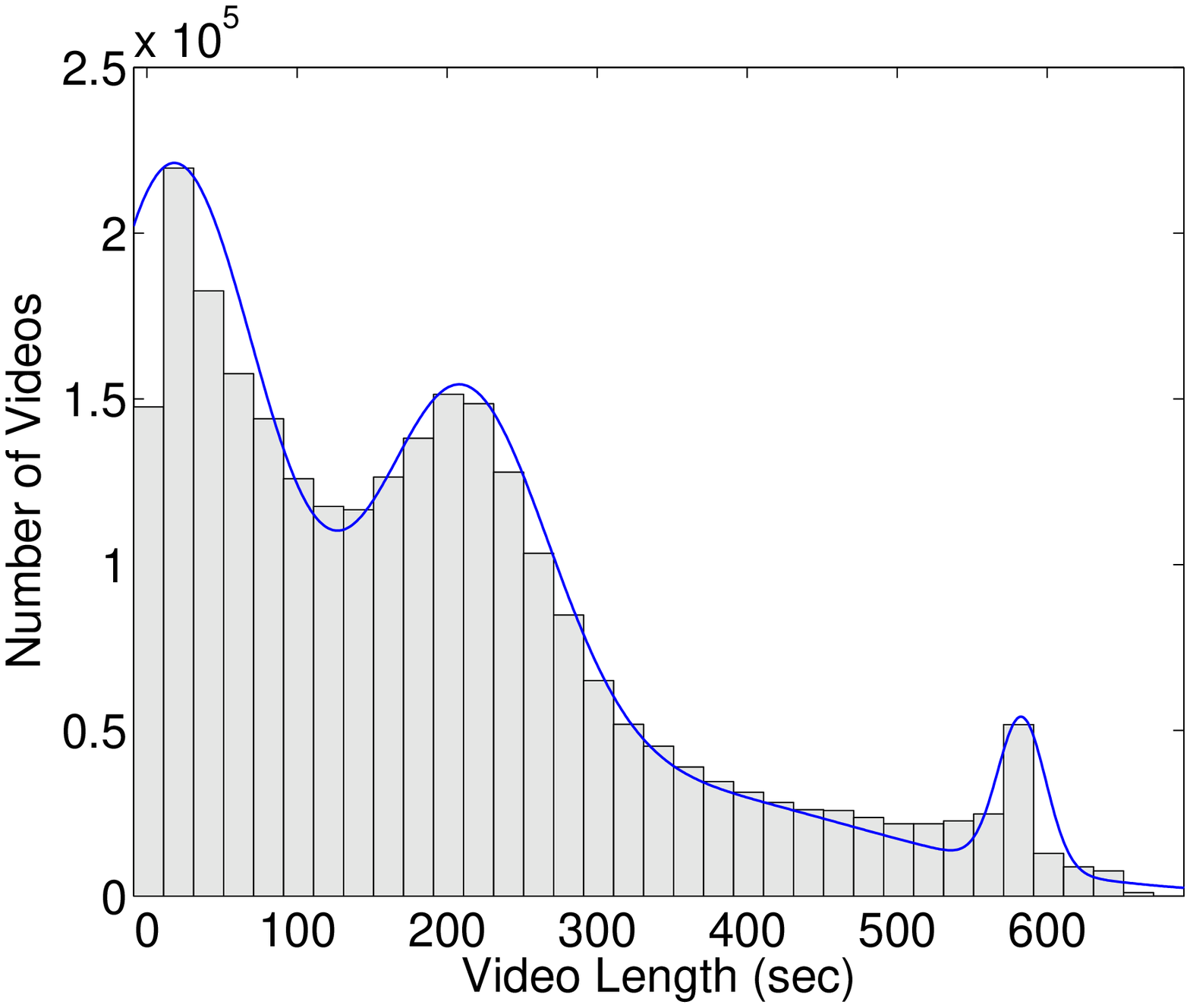}
\caption{Distribution of Video Length} \label{LengthAll}
\end{minipage}
\hspace{0.1cm}
\begin{minipage}{5.5cm}
\centering
\includegraphics[width=\textwidth]{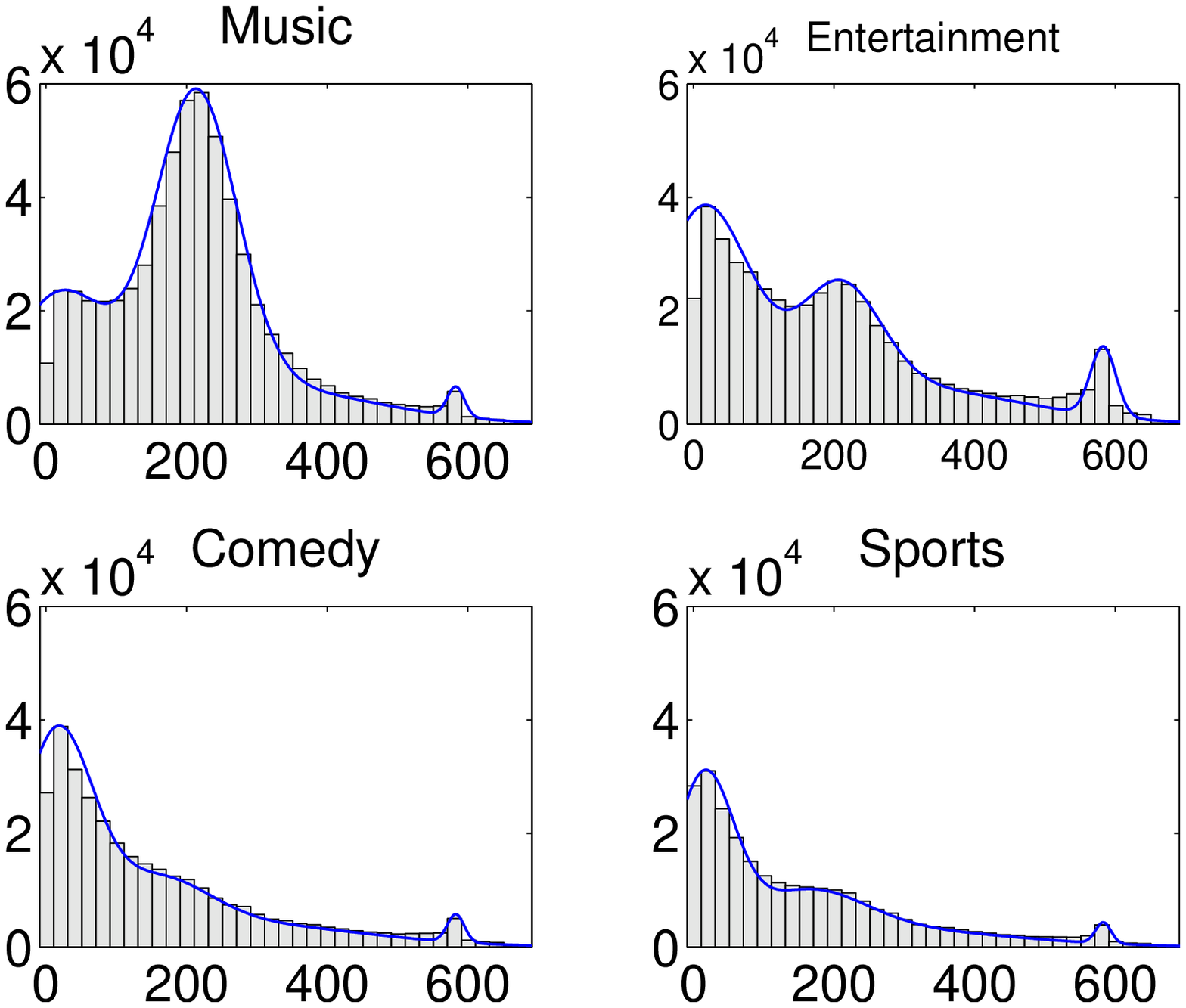}
\caption{Video Length for the 4 Most Popular Categories}
\label{LengthTop4}
\end{minipage}
\hspace{0.1cm}
\begin{minipage}{5.5cm}
\centering
\includegraphics[width=\textwidth]{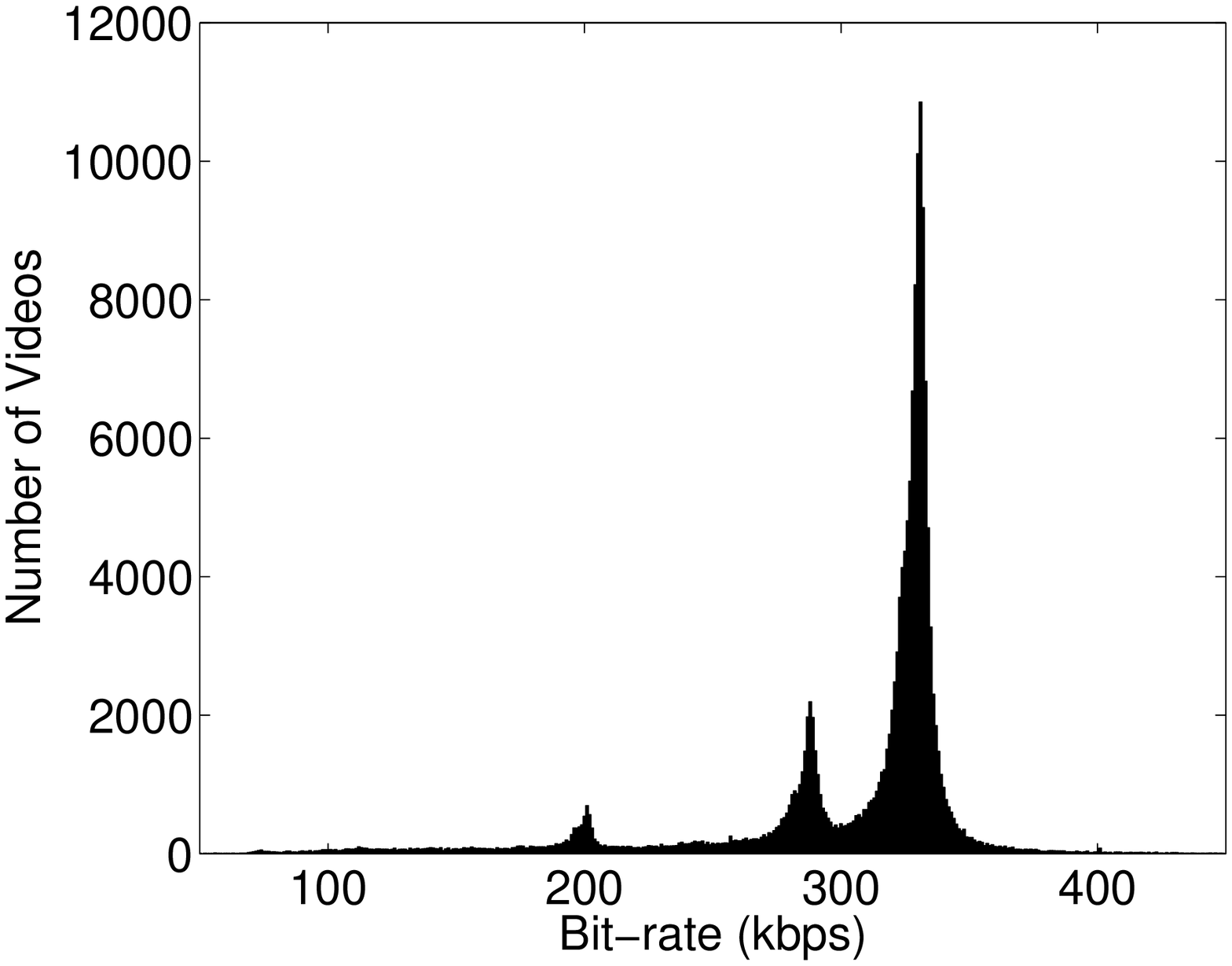}
\caption{Video Bit-rate} \label{Bitrate}
\end{minipage}
\end{figure*}

\subsection{Video Length}\label{SubsecLength}

The length of YouTube videos is the biggest difference from
traditional media content servers. Whereas most traditional servers
contain a small to medium number of long videos, typically 0.5-2
hour movies (e.g. HPLabs Media Server \cite{tang03long}), YouTube is
mostly comprised of videos that are short clips.

In our entire dataset, 97.8\% of the videos' lengths are within 600
seconds, and 99.1\% are within 700 seconds. This is mainly due to
the limit of 10 minutes imposed by YouTube on regular users uploads.
We do find videos longer than this limit though, as the limit was
only established in March, 2006, and also the YouTube Director
Program allows a small group of authorized users to upload videos
longer than 10 minutes \cite{YouTubeBlog}.

Figure \ref{LengthAll} shows the histogram of YouTube videos'
lengths within 700 seconds, which exhibits three peaks. The first
peak is within one minute, and contains more than 20\% of the
videos, which shows that YouTube is primarily a site for very short
videos. The second peak is between 3 and 4 minutes, and contains
about 16.7\% of the videos. This peak is mainly caused by the large
number of videos in the ``Music'' category. Music is the most
popular category for YouTube, and the typical length of a music
video is often within this range, as shown in Figure
\ref{LengthTop4}. The third peak is near the maximum of 10 minutes,
and is caused by the limit on the length of uploaded videos. This
encourages some users to circumvent the length restriction by
dividing long videos into several parts, each being near the limit
of 10 minutes.

We find that the the length histogram can be fit by an aggregate of
four normal distributions, whose parameters are shown in Table
\ref{Para1}. The location parameter $\mu$ determines the mean, the
scale parameter $\sigma$ determines the width, and the ratio $r$
shows the weight of the four curves in the aggregated distribution.
The first three columns in the table correspond to the three peaks
of the distribution, while the last column represents the rest of
the data.

\begin{table}
  \centering
  \begin{tabular}{|c|r|r|r|r|}
    \hline
    \bfseries Parameter & \bfseries Peak 1 & \bfseries Peak 2 &
\bfseries Peak 3 & \bfseries The Rest \\
    \hline
    $\mu$ & 16 & 208 & 583 & 295 \\
    \hline
    $\sigma$ & 62 & 58 & 16 & 172 \\
    \hline
    $r$ & 48.6\% & 26.2\% & 2.7\% & 22.5\% \\
    \hline
  \end{tabular}
  \caption{Parameters for the aggregated normal distribution}
  \label{Para1}
\end{table}

Figure \ref{LengthTop4} shows the video length histograms for the
top four most popular categories. We can see ``Music'' videos have a
very large peak between three and four minutes, and
``Entertainment'' videos have a similar (though smaller) peak. In
comparison, ``Comedy'' and ``Sports'' videos have more videos within
two minutes, probably corresponding to ``highlight'' type clips. We
also used an aggregated normal distribution to get the fits for the
four length distributions.

\subsection{File Size and Bit-rate}\label{SubsecSizeRate}

Using the video IDs from the normal crawl on April 10th 2007 (about
200 thousand videos), we retrieved the file size of nearly 190
thousand videos. In our crawled data, 98.8\% of the videos are less
than 30MB. Not surprisingly, we find that the distribution of video
sizes is very similar to the distribution of video lengths. We
calculate an average video file size to be about 8.4 MBytes.
Considering there are over 42.5 million YouTube videos, the total
disk space required to store all the videos is more than 357
terabytes! Smart storage management is thus quite demanding for such
a ultra-huge and still growing site, which we will discuss in more
detail in Section \ref{SecImpli}.

We found that 87.6\% of the videos we crawled contained FLV
meta-data specifying the video's bit-rate in the beginning of the
file, indicating that they are Constant-Bit-Rate (CBR). For the rest
of the videos that do not contain this meta-data (probably
Variable-Bit-Rate, or VBR, videos), we calculate an average bit-rate
from the file size and its length.

In Figure \ref{Bitrate}, the videos' bit-rate has three clear peaks.
Most videos have a bit-rate around 330 kbps, with two other
peaks at around 285 kbps and 200 kbps. This implies that YouTube
videos have a moderate bit-rate that balances the quality and the
bandwidth.

\begin{figure*}
\center
\begin{minipage}{5.5cm}
\centering
\includegraphics[width=\textwidth]{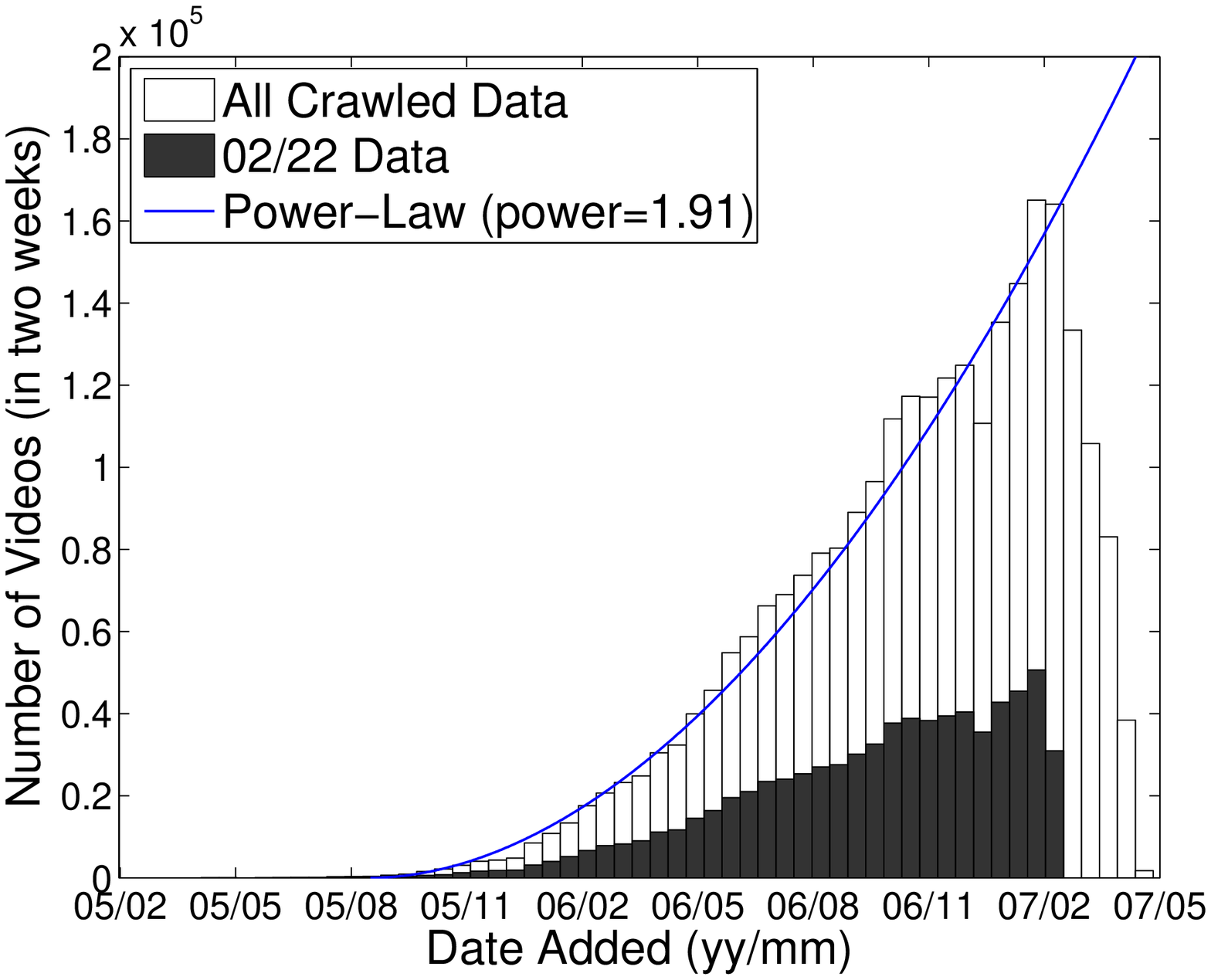}
\caption{Date Added} \label{DateAdd}
\end{minipage}
\hspace{0.1cm}
\begin{minipage}{5.5cm}
\centering
\includegraphics[width=\textwidth]{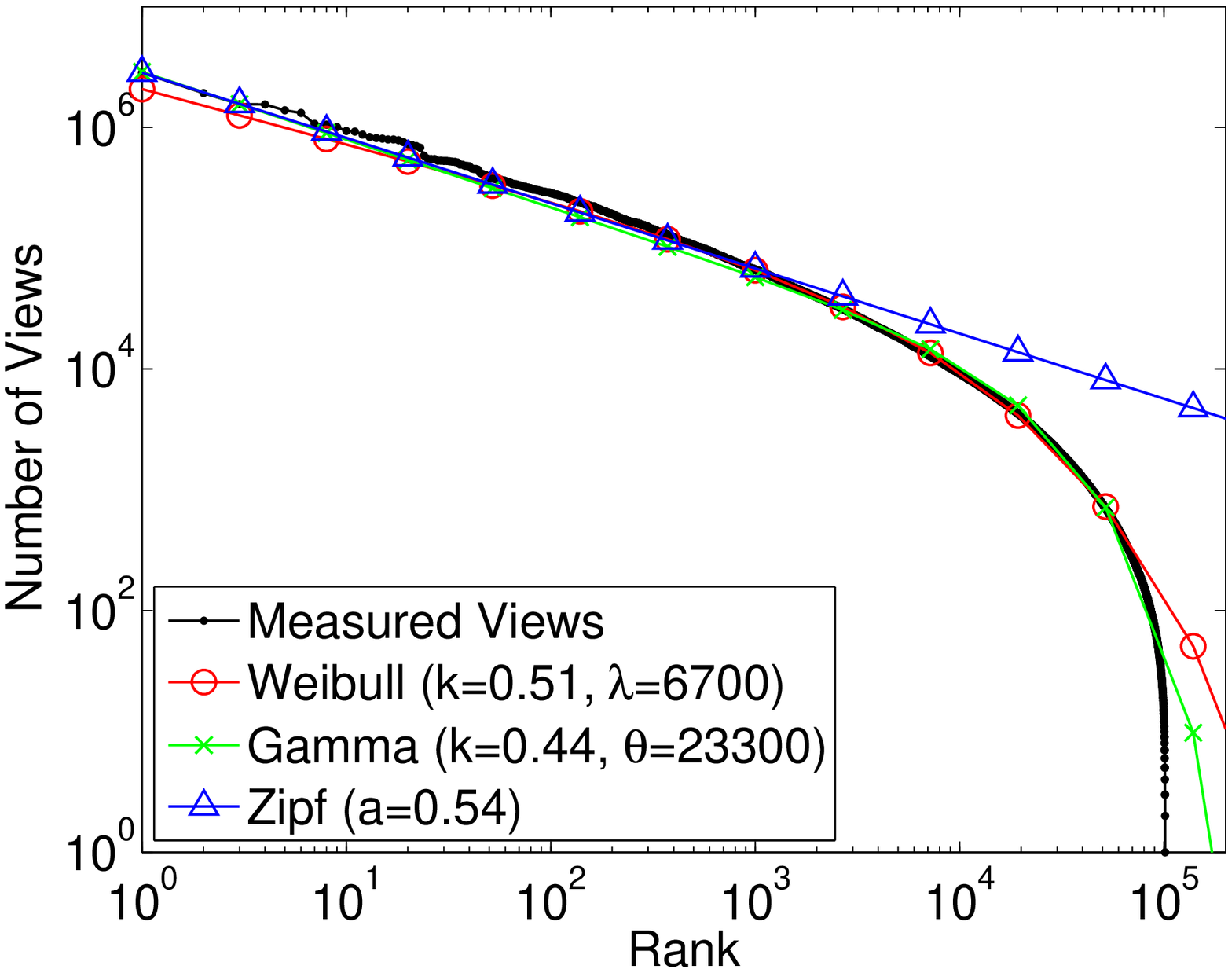}
\caption{Rank of Views} \label{ViewsRank}
\end{minipage}
\hspace{0.1cm}
\begin{minipage}{5.5cm}
\centering
\includegraphics[width=\textwidth]{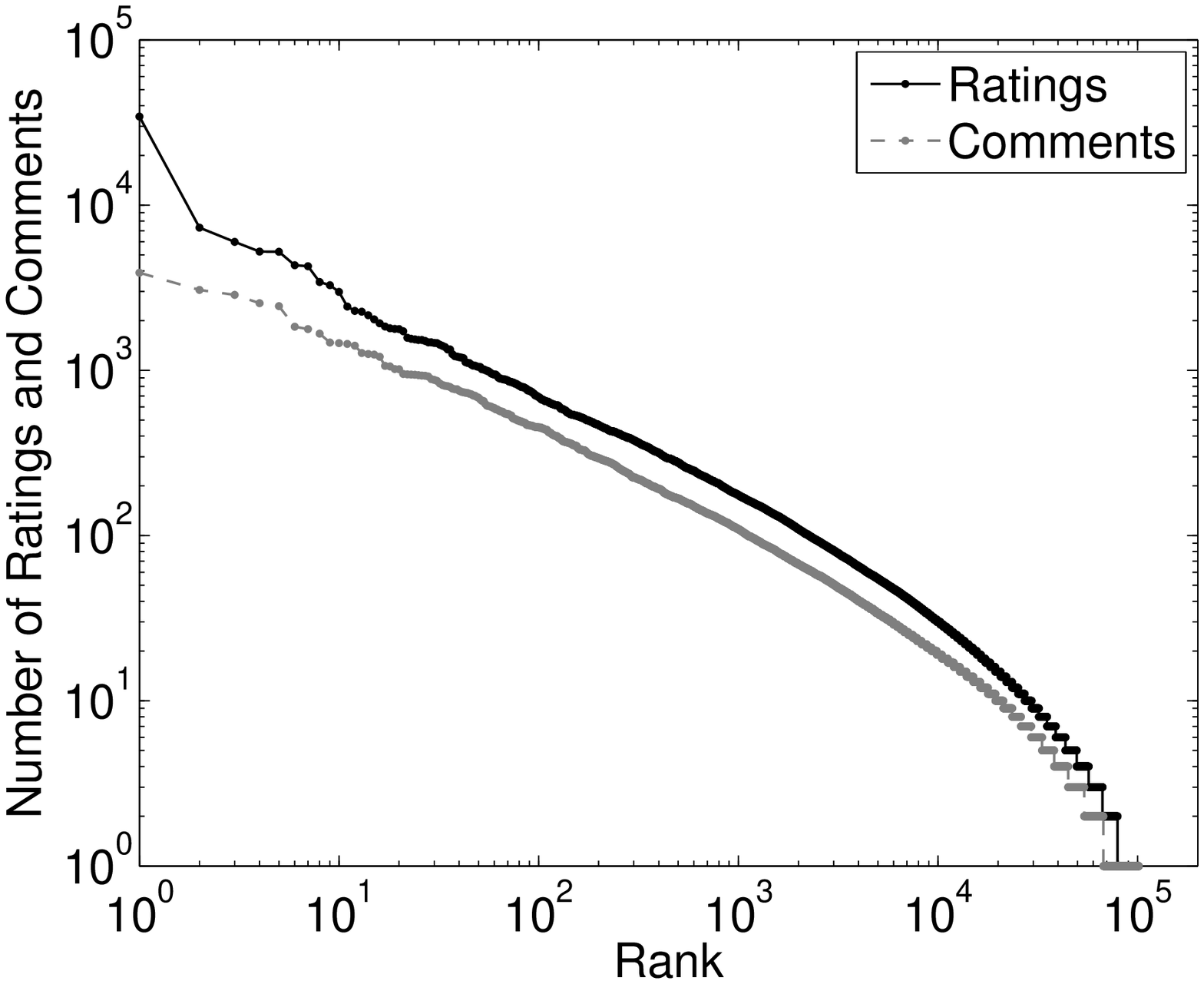}
\caption{Rank of Ratings and Comments} \label{RatingsCommentsRank}
\end{minipage}
\end{figure*}

\subsection{Date Added -- Growth Trend of Uploading}
\label{SubsecUpload}

During our crawl we record the date that each video was uploaded, so
we can study the growth trend of YouTube. Figure \ref{DateAdd} shows
the number of new videos added every two weeks in our entire crawled
dataset.

February 15th, 2005 is the day that YouTube was established. Our
first crawl was on February 22nd, 2007, thus we can get the early
videos only if they are still very popular videos or are linked to
by other videos we crawled. We can see there is a slow start, the
earliest video we crawled was uploaded on April 27th, 2005. After 6
months from YouTube's establishment, the number of uploaded videos
increases steeply. We use a power law curve to fit this trend.

In the dataset we collected, the number of uploaded videos decreases
linearly and steeply starting in March, 2007. However, this does
not imply that the uploading rate of YouTube videos has suddenly
decreased. The reason is that many recently uploaded videos have not
been so popular, and are probably not listed in other videos related
videos' list. Since few videos have linked to those new videos, they
are not likely to be found by our crawler. Nevertheless, as those
videos become popular or get linked to by others, our crawler may
find them and get their information. Comparing the entire dataset to
the first and largest dataset, which was crawled on February 22nd,
we also see the same trend.

\subsection{Views, Ratings -- User Access Pattern}
\label{SubsecAccess}

The number of views a video has had is the most important
characteristic we measured, as it reflects the popularity and access
patterns of the videos. Because this property is changing over time,
we cannot use the entire dataset that combines all the data
together. Therefore we use a single dataset from April 3rd, 2007,
containing more than 100 thousand videos, which is considered to be
relatively static.

Figure \ref{ViewsRank} shows the number of views as a function of
the rank of the video by its number of views. Though the plot has a
long tail on the linear scale, it does NOT follow a Zipf
distribution, which should be a straight line on a log-log scale.
This is consistent with some previous observations
\cite{yu06understanding, tang03long, almeida01edu,
acharya00characterizing} that also found that video accesses on a
media server does not follow Zipf's law. We can see in the figure,
the beginning of the curve is linear on a log-log scale, but the
tail (after the $2\times10^3$ video) decreases tremendously,
indicating there are not so many less popular videos as Zipf's law
predicts. This result seems consistent with some results
\cite{yu06understanding}, but differs from others \cite{tang03long,
almeida01edu, acharya00characterizing} in which the curve is skewed
from linear from beginning to end. Their results indicate that the
popular videos are also not as popular as Zipf's law predicts, which
is not the case in our experiment.

To fit the skewed curve, some use a generalized Zipf-like
distribution \cite{tang03long}, while others use a concatenation of
two Zipf-like distributions \cite{acharya00characterizing}. Because
our curve is different, we attempted to use three different
distributions: Weibull, Gamma and Zipf. We find that
Weibull and Gamma distributions both fit better than Zipf, due to
the drop-off in the tail (in log-log scale) that they have.

Figure \ref{RatingsCommentsRank} plots the number of ratings against
the rank of the video by the number of ratings,
and similarly for the number of comments. The two both have the same
distribution, and are very similar to the plot of the number of
views in Figure \ref{ViewsRank}, yet the tails of the two do not
drop so quickly compared to that of number of views.

\begin{figure*}
\center
\begin{minipage}{5.5cm}
\centering
\includegraphics[width=\textwidth]{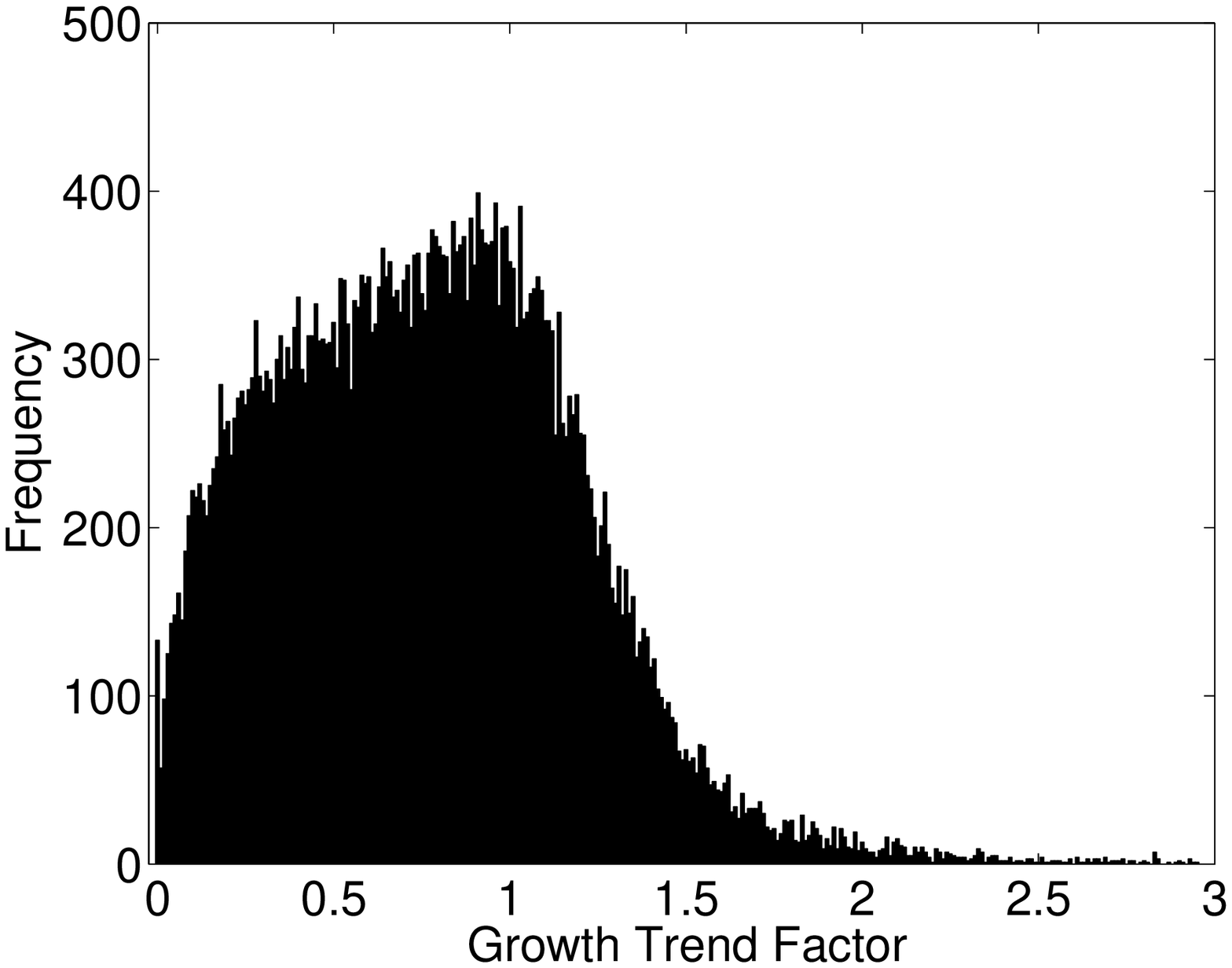}
\caption{Growth Trend Factor} \label{GrowthTrend}
\end{minipage}
\hspace{0.1cm}
\begin{minipage}{5.5cm}
\centering
\includegraphics[width=\textwidth]{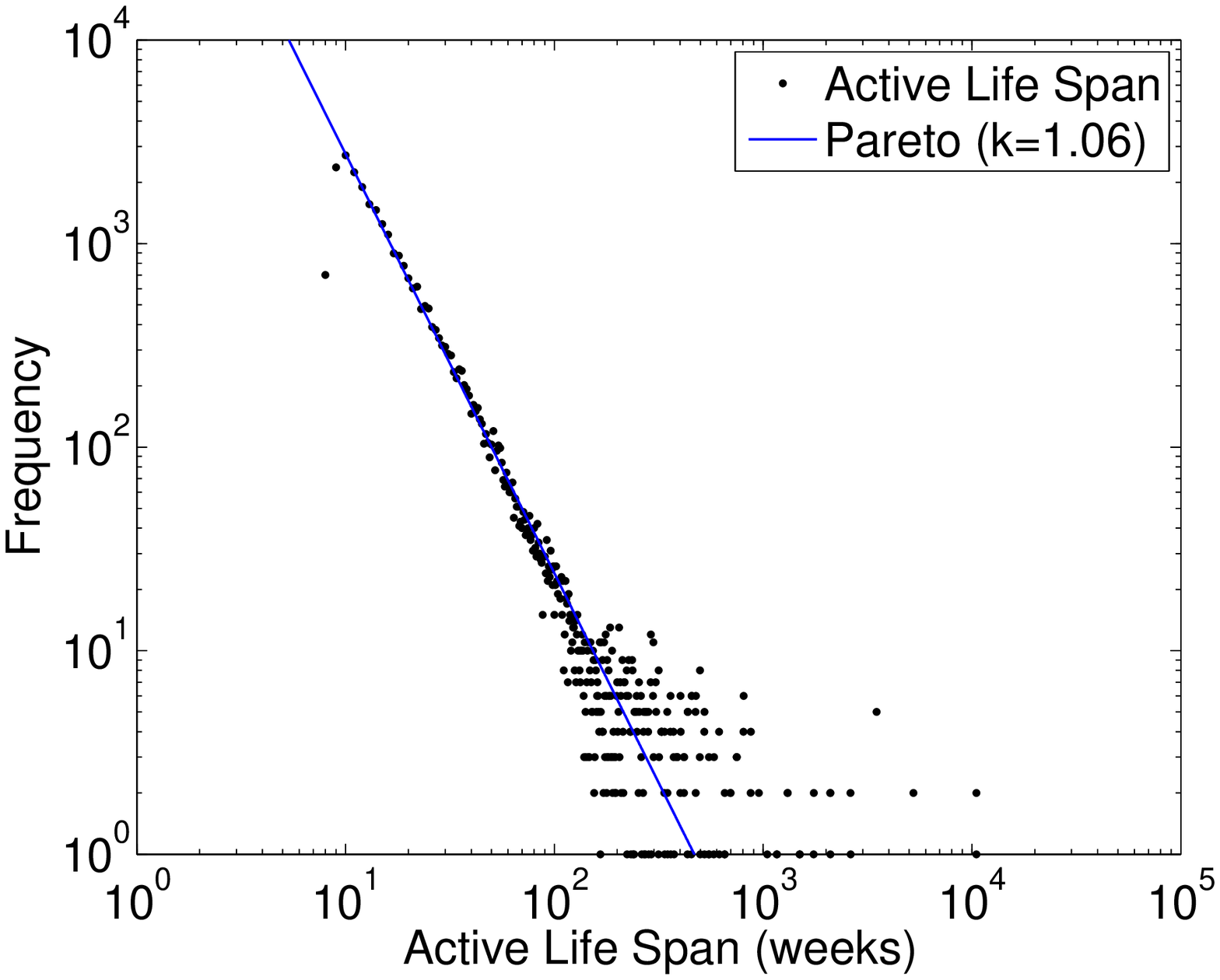}
\caption{Estimated Active Life Span (t=10\%)} \label{LifeSpanRank}
\end{minipage}
\hspace{0.1cm}
\begin{minipage}{5.5cm}
\centering
\includegraphics[width=\textwidth]{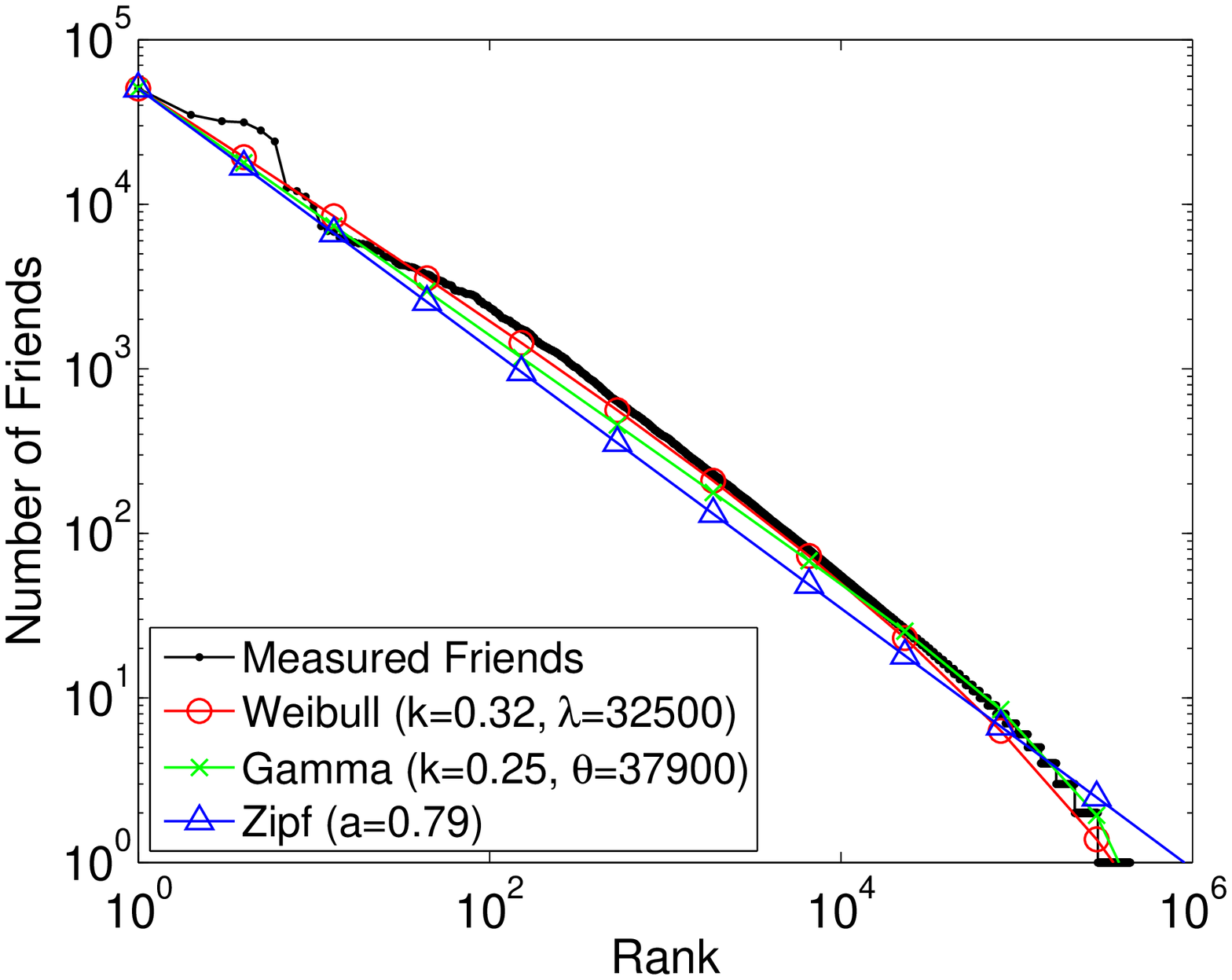}
\caption{Rank of User Friends} \label{UserFriends}
\end{minipage}
\end{figure*}

\subsection{Growth Trend of Number of Views and Active Life Span}
\label{SubsecGrowth}

Comparing the popularity of YouTube videos, we find that some are
very popular (their number of views grows very fast), while others
are not. Also, after a certain period of time, some videos are
almost never watched.

Starting on March 5th, 2007, we updated the number of views
statistic of relatively new videos (uploaded after February 15th,
2007) every week for seven weeks. To be sure the growth trend will
be properly modelled, we eliminate any videos that have been removed
and so do not have the full seven data points, resulting in a
dataset size totaling approximately 43 thousand videos.

We have found that the growth trend can be modeled better by a power
law than a linear fit. Therefore, a video's growth
trend can be increasing (if the power is greater than 1), growing
relatively constantly (power near 1), or slowing in
growth (power less than 1). The trend depends on the exponent
factor used in the power law, which we call the growth trend factor
$p$. We define the views count after $x$ weeks as
\begin{equation}
\label{growthfactor} v(x) = v_0 \times \frac{(x+\mu)^p}{\mu^p}
\end{equation}
where $\mu$ is the number of weeks before March 5th that the video
has been uploaded, $x$ indicates the week of the crawled data (from
0 to 6), and $v_0$ is the number of views the video had on March
5th.

We modelled the 43 thousand videos using equation \ref{growthfactor}
to get the distribution of growth trend factors $p$, which is shown
in Figure \ref{GrowthTrend}. Over 70\% of the videos have a growth
trend factor that is less than 1, indicating that most videos grow
in popularity more slowly as time passes.

Since YouTube has no policy on removing videos after a period of
time or when their popularity declines, the life span of a YouTube
video is almost infinite. However, when the video's popularity grows
more and more slowly, the popularity growth curve will become
horizontal. Since it will almost stop growing after some
time, we will define that as the video's active life span. From this
active life span, we can extract some characteristics of the
temporal locality of videos.

If a video's number of views increases by a factor less than $t$
from the previous week, we define the video's active life span to be
over. We prefer this relative comparison to an absolute comparison,
since we are only concerned with the shape of the curve instead of
the scale.

For each video that has a growth trend factor $p$ less than 1, we
can compute its active life span $l$ from
\begin{equation}
\frac{v(l)}{v(l-1)} - 1 = t \nonumber
\end{equation}
which can be solved for the active life span
\begin{equation}
\label{lifespan} l = \frac{1}{\sqrt[p]{1+t}-1}+1-\mu
\end{equation}
Thus we see that the active life span is dependent on the growth
trend factor $p$ and the number of weeks the video has been on
YouTube, but does not depend on the number of views the video had at
the start of the experiment.

Figure \ref{LifeSpanRank} shows the probability density function
(PDF) for the active life span of the approximately 30 thousand
videos (with $p$ less than 1), for a life span factor of $t=10\%$.
The solid line is the Pareto distribution fit to the data, which
fits very well, and results in a parameter $k$ of 1.06. From looking
at multiple fits with various values of $t$, we find that they all
result in the same parameter $k$, the only difference is the
location of the line.

Since we do not have the server logs of YouTube, we cannot
accurately measure the characteristic of temporal locality, which
would show whether recently accessed videos are likely to be
accessed in the near future. However, the active life span gives us
another way to view the temporal locality of YouTube videos. Figure
\ref{LifeSpanRank} implies that most videos have a short active life
span, which means the videos have been watched frequently in a short
span of time. Then, after the video's active life span is complete,
fewer and fewer people will access them.

This characteristic has good implications for web caching and server
storage. We can design a predictor to predict the active life span
using our active life span model from equation \ref{lifespan}. The
predictor can help a proxy or server to make more intelligent
decisions, such as when to drop a video from the cache. We will
discuss this in more detail in Section \ref{SecImpli}.


\section{The Social Network in YouTube}\label{SecSocial}

YouTube is a prominent social media application: there are
communities and groups in YouTube, there are statistics and awards
for videos and personal channels. Videos are no longer independent
from each other, and neither are the users. It is therefore
important to understand the social network characteristics of
YouTube. We next examine the social network among YouTube users and
videos, which is a very unique and interesting aspect of this kind
of video sharing sites, as compared to traditional media services.

\subsection{User Friends and Upload}\label{SubsecUser}

We have examined the relations among the YouTube users. From the
crawl of user information we performed on May 28th 2007, we can
extract two characteristics of YouTube users: the number of friends
and the number of uploaded videos. We did this for the more than 1
million users found by our crawler in all the crawls performed
before this one.

Figure \ref{UserFriends} shows the number of friends each user has,
compared with the rank of the user by the number of friends.
Compared with previous plots, it is much closer to linear on
a log-log scale, though we still use the same three distributions
to get the best fit. Interestingly, in over 1 million users' data,
we found that 58\% of the user's have no friends. We believe that
this is partially because YouTube is still quite young, with more
connections to be established between its users.

We also plotted the number of uploaded videos each user has,
compared with the rank of the user by number of uploads. It is very
similar to the previous plots of the number of views and friends,
and so we omit it for brevity.

\subsection{Small-World Networks}

\begin{figure*}
\center
\begin{minipage}{5.5cm}
\centering
\includegraphics[width=\textwidth]{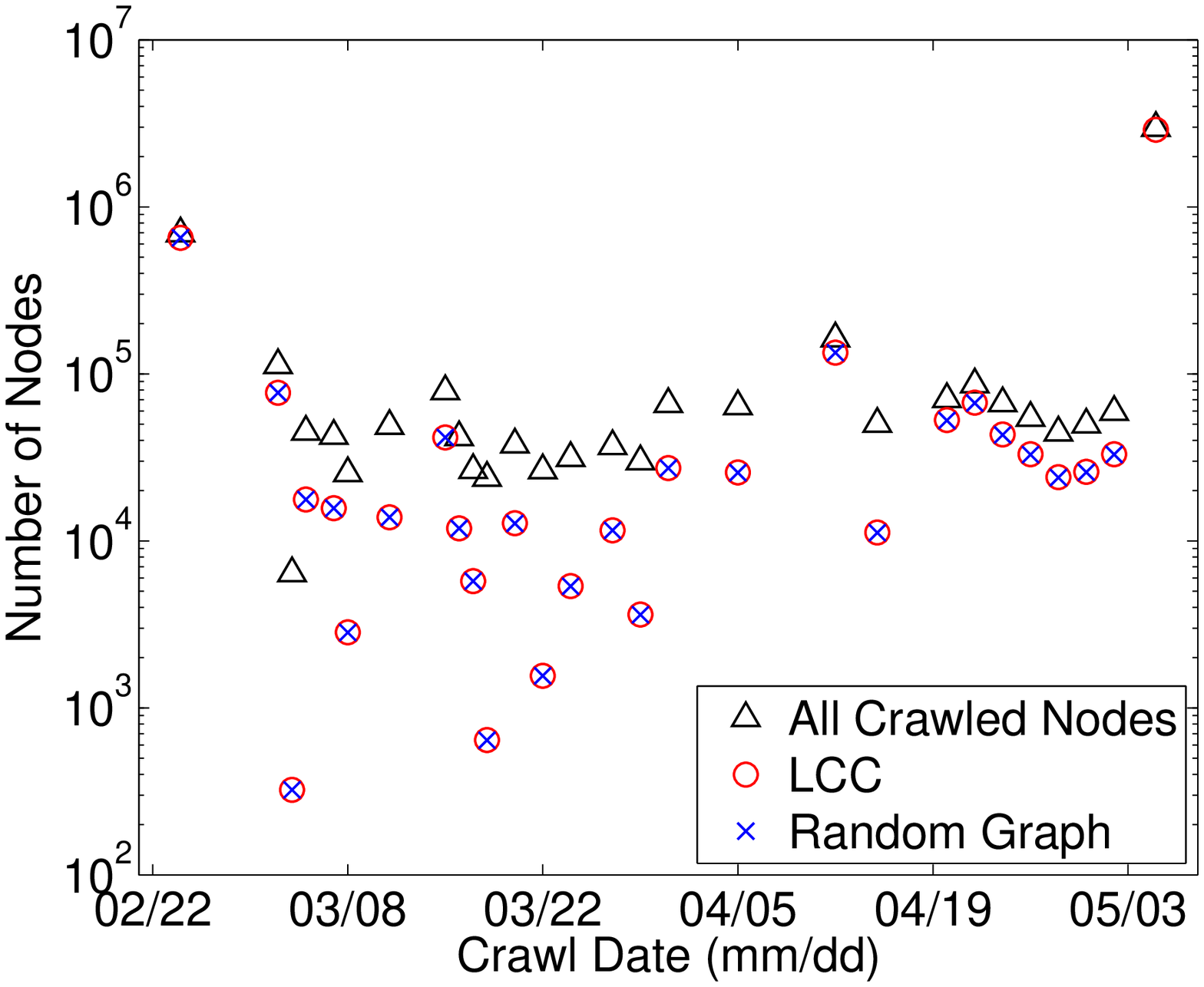}
\caption{Dataset Size} \label{dataset}
\end{minipage}
\hspace{0.1cm}
\begin{minipage}{5.5cm}
\centering
\includegraphics[width=\textwidth]{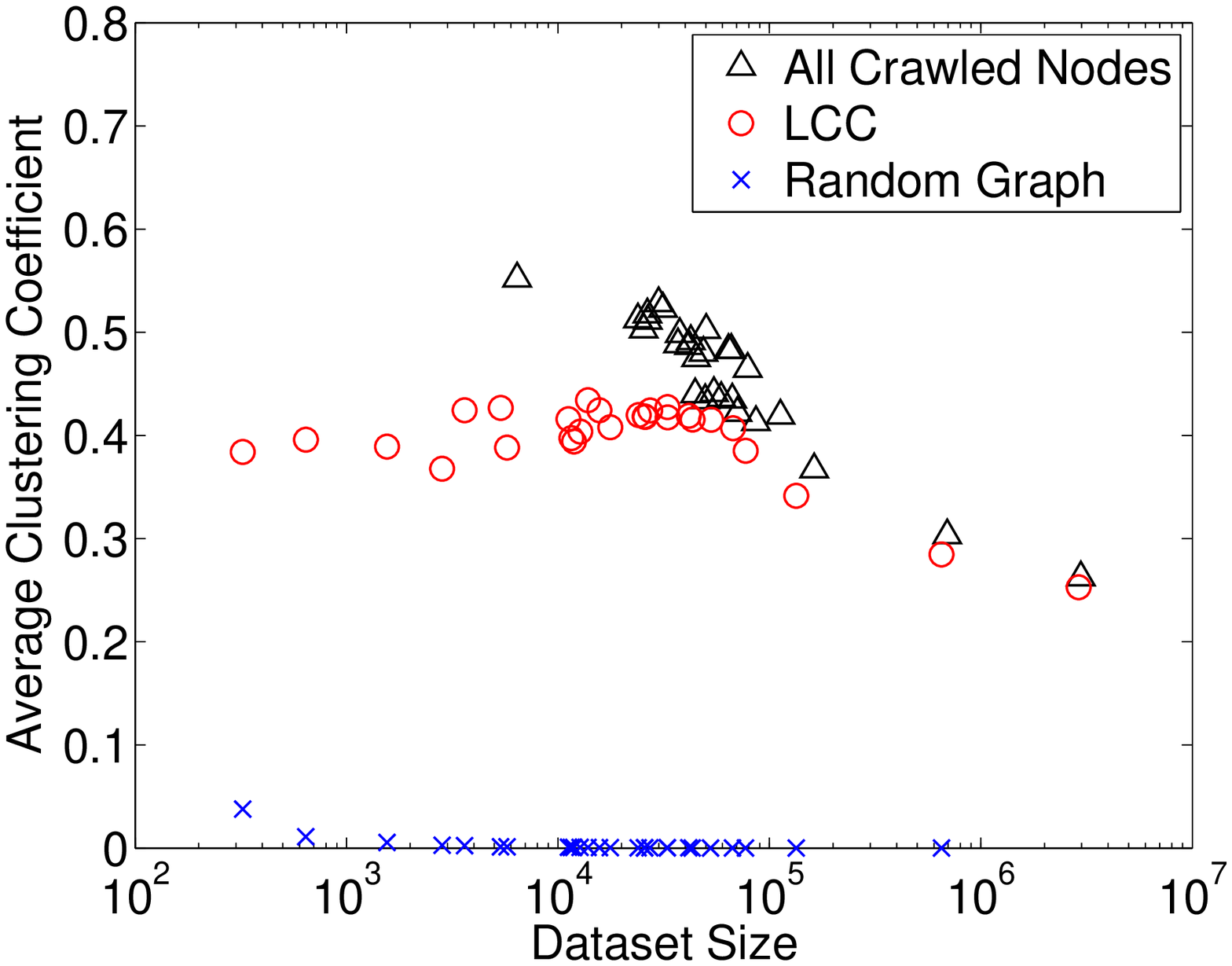}
\caption{Clustering Coefficient} \label{cluster}
\end{minipage}
\hspace{0.1cm}
\begin{minipage}{5.5cm}
\centering
\includegraphics[width=\textwidth]{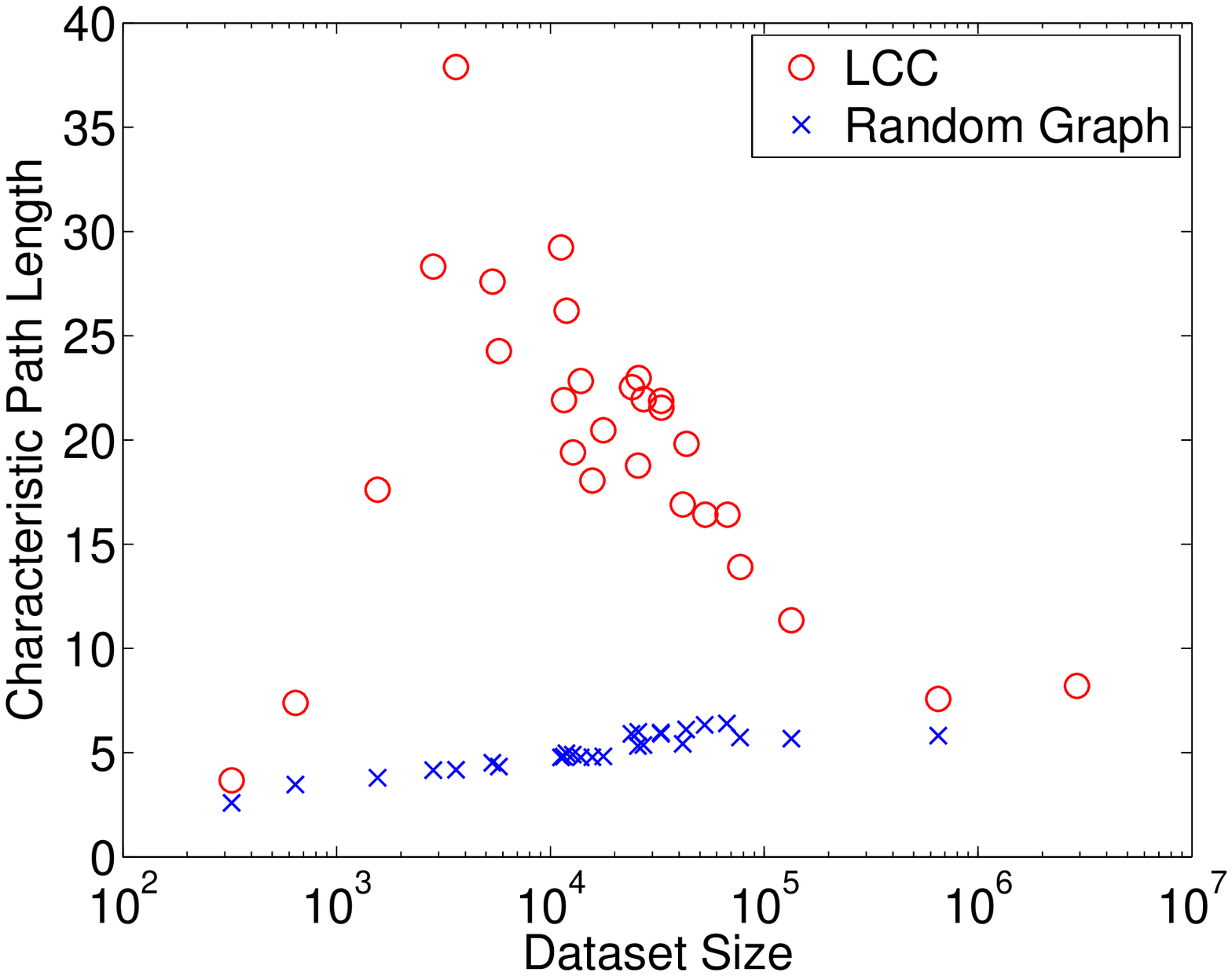}
\caption{Characteristic Path Length} \label{diameter}
\end{minipage}
\end{figure*}

Small-world network phenomenon is probably the most interesting
characteristic for social networks. It has been found in various
real-world situations: URL links in the Web \cite{albert1999dww},
Gnutella's search overlay topology \cite{liu2004mam}, and Freenet's
file distribution network \cite{hong2001perf}.

The concept of a small-world  was first introduced by Milgram
\cite{milgram1967swp} to refer to the the principle that people are
linked to all others by short chains of acquaintances (popularly
known as \emph{six degrees of separation}). This formulation was
used by Watts and Strogatz to describe networks that are neither
completely random, nor completely regular, but possess
characteristics of both \cite{watts1998cds}. They
introduce a measure of one of these characteristics, the
cliquishness of a typical neighborhood, as the \emph{clustering
coefficient} of the graph. They define a small-world graph as one in
which the clustering coefficient is still large, as in regular
graphs, but the measure of the average distance between nodes (the
\emph{characteristic path length}) is small, as in random graphs.

Given the network as a graph $G = (V,E)$, the clustering coefficient
$C_i$ of a node $i \in V$ is the proportion of all the possible
edges between neighbors of the node that actually exist in the
graph. The clustering coefficient of the graph $C(G)$ is then the
average of the clustering coefficients of all nodes in the graph.
The characteristic path length $d_i$ of a node $i \in V$ is the
average of the minimum number of hops it takes to reach all other
nodes in $V$ from node $i$. The characteristic path length of the
graph $D(G)$ is then the average of the characteristic path lengths
of all nodes in the graph.

\subsection{The Small-World in YouTube}
\label{SubsecSmallWorld}

We measured the graph topology for all the YouTube data gathered,
by using the related links in YouTube pages to form directed edges
in a video graph for each dataset. Videos that have no outgoing or
no incoming links are removed from the analysis. In addition, a
combined dataset consisting of all the crawled data integrated into
one set is also created. Since not all of YouTube is crawled, the
resulting graphs are not strongly connected, making it difficult to
calculate the characteristic path length. Therefore, we also use the
Largest strongly Connected Component (LCC) of each graph for the
measurements. Every crawled dataset therefore results in 2 graphs,
plus 2 more graphs for the combined dataset.

For comparison, we also generate random graphs that are strongly
connected. Each of the random graphs has the same number of nodes
and average node degree of the strongly connected component of the
crawled dataset, and is also limited to a maximum node out-degree of
20, similar to the crawled datasets. The only exception is the
combined dataset of all the crawled data, which was too large to
generate a comparable random graph for.

Some graphs use the dataset size for the x-axis values, so that we
can see trends as the dataset size increases. This is very
informative, as we are not mapping the entire YouTube website, but
only a portion of it. Therefore, some extrapolation as the dataset
size increases will be needed to draw insights into the graph formed
by all of the YouTube videos.

Figure \ref{dataset} shows the dataset sizes and the date they were
created on. It also has the strongly connected component size and
the random graph size, both of which are very close to the total
dataset size for the larger datasets. The combined dataset is also
shown, and is given the most recent date. By far the largest crawled
dataset is the first one, crawled on Feb 22.

Figure \ref{cluster} shows the average clustering coefficient for
the entire graph, as a function of the size of the dataset. The
clustering coefficient is quite high in most cases, especially in
comparison to the random graphs. There is a noticeable drop in the
clustering coefficient for the largest datasets, showing that there
is some inverse dependence on the size of the graph, which is common
for some small-world networks \cite{ravasz2003hoc}.

Figure \ref{diameter} shows the characteristic path length for each
of the datasets' graphs. There are two factors influencing the shape
of the graph. As the dataset size increases, the maximum possible
diameter increases, which is seen in the smallest datasets. Once the
dataset reaches a size of a few thousand nodes, the diameter starts
to decrease as the small-world nature of the graph becomes evident.
For the largest datasets, the average diameter is only slightly
larger than the diameter of a random graph, which is quite good
considering the still large clustering coefficient of these
datasets.

The network formed by YouTube's related videos list has definite
small-world characteristics. The clustering coefficient is very
large compared to a similar sized random graph, while the
characteristic path length of the larger datasets are approaching
the short path lengths measured in the random graphs. This finding
is expected, due to the user-generated nature of the tags, title and
description of the videos that is used by YouTube to find related
ones.

These results are similar to other real-world user-generated graphs
that exist, yet their parameters can be quite different. For
example, the graph formed by URL links in the world wide web
exhibits a much longer characteristic path length of 18.59
\cite{albert1999dww}. This could possibly be due to the larger
number of nodes ($8 \times 10^8$ in the web), but it may also
indicate that the YouTube network of videos is a much closer group.


\begin{figure*}
\center
\begin{minipage}{5.5cm}
\centering
\includegraphics[width=\textwidth]{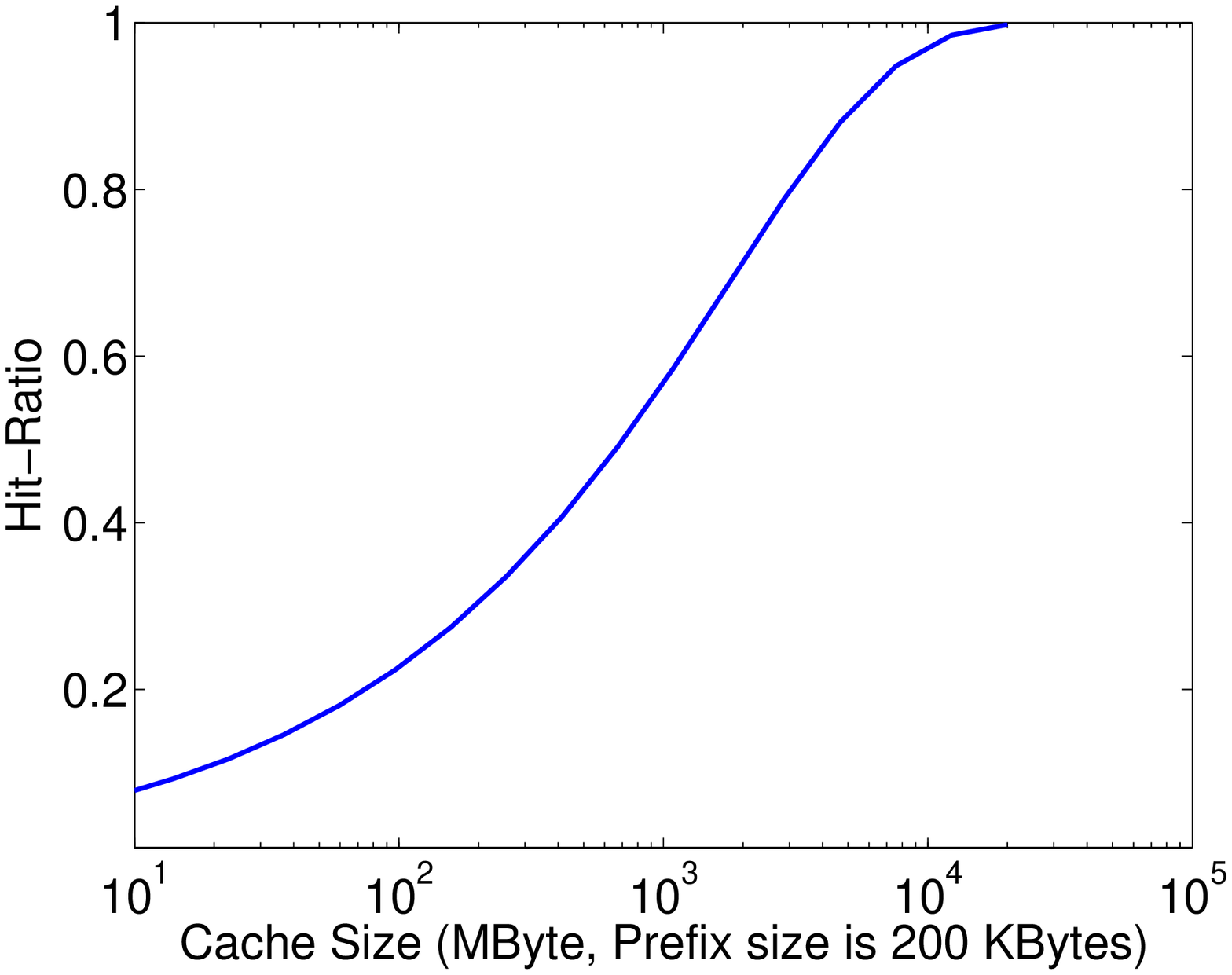}
\caption{Cache Size to Hit-ratio} \label{hitratio}
\end{minipage}
\hspace{0.1cm}
\begin{minipage}{5.5cm}
\centering
\includegraphics[width=\textwidth]{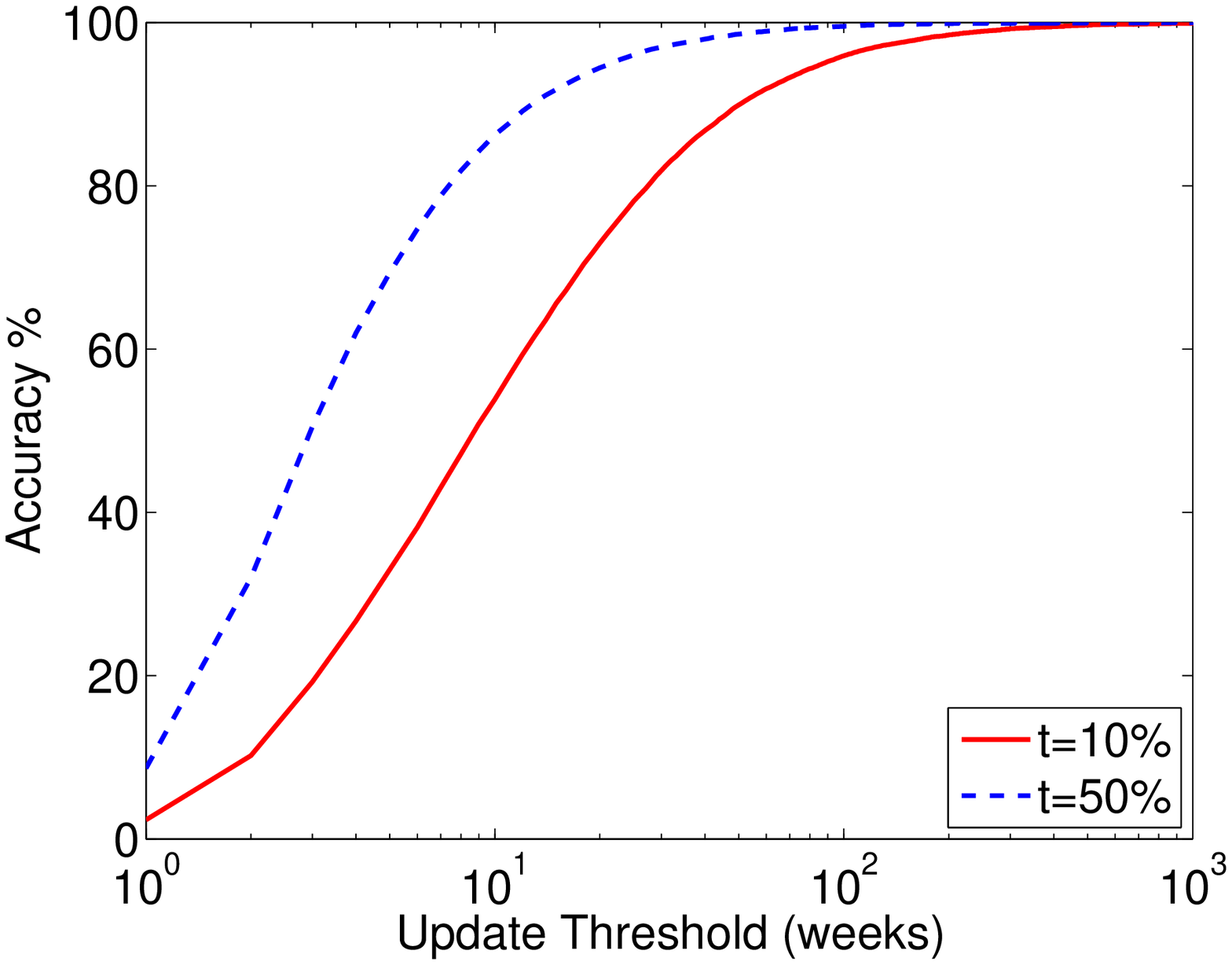}
\caption{Accuracy to Update Threshold} \label{prediction}
\end{minipage}
\hspace{0.1cm}
\begin{minipage}{5.5cm}
\centering
\includegraphics[width=\textwidth]{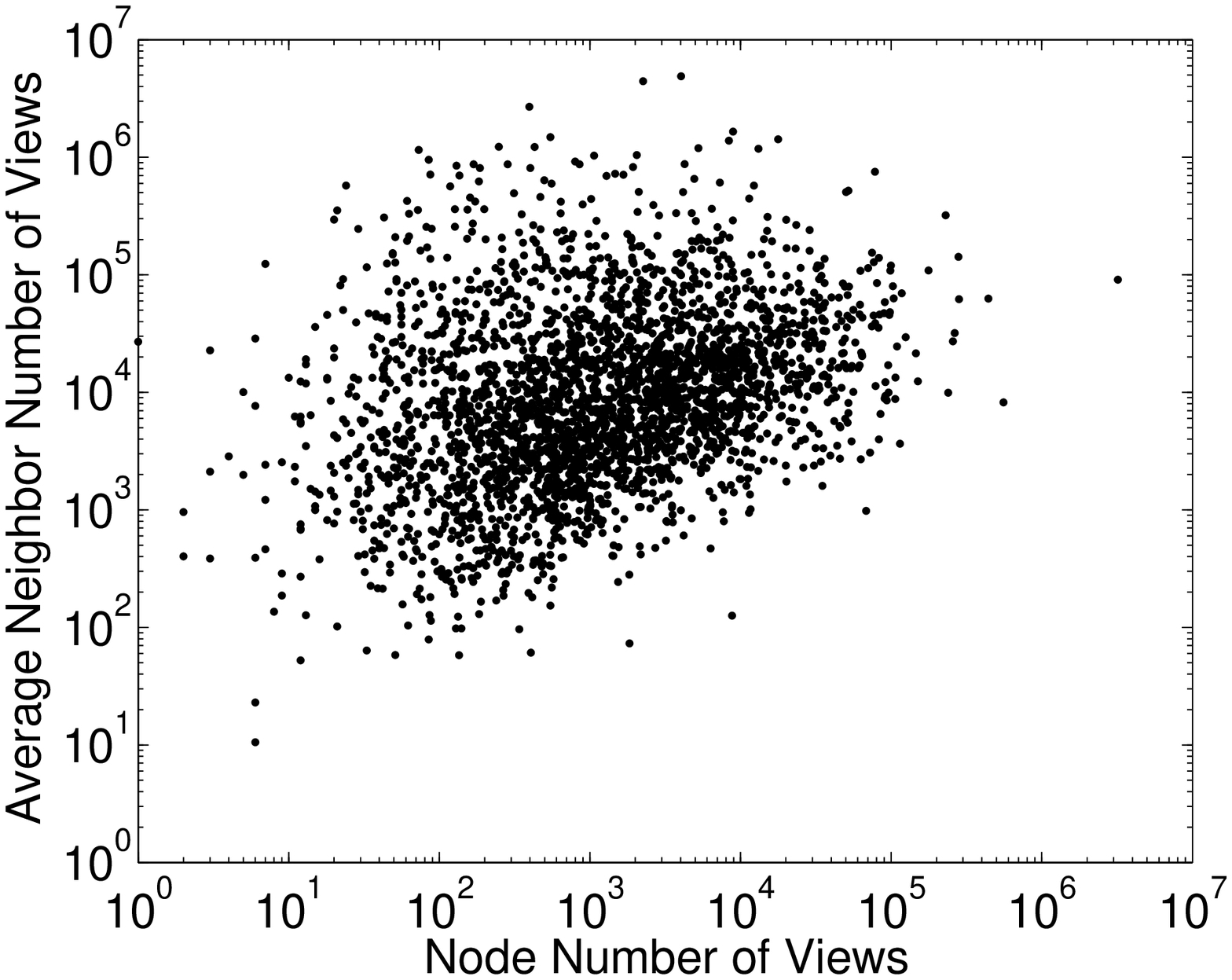}
\caption{Views to Neighbor Views Correlation} \label{nviews}
\end{minipage}
\end{figure*}

\section{Further Discussions}\label{SecImpli}

A very recent study shows that YouTube alone has comprised
approximately 20\% of all HTTP traffic, or nearly 10\% of all
traffic on the Internet, with a nearly 20\% growth rate per month
\cite{YouTubeBandwidth, corbett06peering}. Assuming the
network traffic cost is \$10/Mbps, the estimated YouTube transit
expenses is currently more than \$2 million per month. This high and
rising expense for network
traffic is probably one of the reasons YouTube was sold to Google.

According to Alexa \cite{ALEXA}, the current speed of YouTube has
become ``Very Slow'' and is considered slower than 81\% of the
surveyed sites. This situation is only getting worse. Scalability is
no doubt the biggest challenge that YouTube faces, particularly
considering that websites such as YouTube survive by attracting more
users. In this section, we briefly discuss the implications of our
measurement results toward improving the scalability of YouTube.

\subsection{Implications on Proxy Caching and Storage Management}
\label{SubsecCache}

Caching frequently used data at proxies close to clients is an
effective way to save backbone bandwidth and prevent users from
experiencing excessive access delays. Numerous algorithms have been
developed for caching web objects or streaming videos. While we
believe that YouTube will benefit from proxy caching
\cite{liu2004pcm}, three distinct features call for novel cache
designs. First, the number of YouTube videos (42.5 million
\cite{tang03long}) is orders of magnitude higher than that of
traditional video streams (e.g. HPC: 2999, HPL: 412
\cite{VideoNumber}). The size of YouTube videos is also much smaller
than a traditional video (98.8\% are less than 30MB in YouTube
versus a typical MPEG-1 movie of 700MB). Finally, the view
frequencies of YouTube videos do not well fit a Zipf distribution,
which has important implications on web caching \cite{breslau99web}.

Considering these factors, full-object caching for web or segment
caching for streaming video are not practical solutions for YouTube.
Prefix caching \cite{sen99proxy} is probably the best choice. Assume
for each video, the proxy will cache a 5 second initial
clip, i.e. about 200KB of the video. Given the Gamma distribution of
view frequency suggested by our measurements, we plot the
hit-ratio as a function of the cache size in Figure \ref{hitratio},
assuming that the cache space is devoted only to the most popular
videos. To achieve a 60\% hit-ratio, the proxy would require about 1
GByte of disk space for the current YouTube video repository, and
nearly 8 GByte for a 95\% hit-ratio. Such demand on disk space is
acceptable for today's proxy servers.

Given the constant evolution of YouTube's video repository, a
remaining critical issue is when to release the space for a cached
prefix. We found in Section \ref{SubsecGrowth} that the active life
span of YouTube videos follows a Pareto distribution, implying that
most videos are popular during a relatively short span of time.
Therefore, a predictor can be developed to forecast the active life
span of a video. With the predictor, the proxy can decide which
videos have already passed their life span, and replace it if the
cache space is insufficient.

The life span predictor can also facilitate disk space management on
the YouTube server. Currently, videos on a YouTube server will not
be removed by the operator unless they violate the terms of service.
With a daily 65,000 new videos introduced, the server storage will
soon become a problem. A hierarchical storage structure can be built
with videos passing their active life span being moved to slower and
cheaper storage media. From our 30 thousand videos dataset (Section
\ref{SubsecGrowth}), we calculate the predictor accuracy from the
number of videos that have an active life span (according to
equation \ref{lifespan}) less than an update threshold divided by
the total number of videos, which is plotted in Figure
\ref{prediction}. This result facilitates the determination of an
update threshold for the predictor with a given accuracy.

The cache efficiency can be further improved by exploring the
small-world characteristic of the related video links (see Section
\ref{SubsecSmallWorld}). That is, if a group of videos have a tight
relation, then a user is likely to watch another video in the group
after finishing the first one. This expectation is confirmed by
Figure \ref{nviews}, which shows a clear correlation between the
number of views and the average of the neighbors' number of views.
Once a video is played and cached, the prefixes of its directly
related videos can also be prefetched and cached, if the cache space
allows. We have evaluated the effectiveness of this prefetching
strategy, which shows that the resultant hit-ratio is almost the
same as that of always caching the most popular videos, and yet its
communication overhead is significantly lower because it does not
have to keep track of the most popular videos list.

\subsection{Can Peer-to-Peer Save YouTube?}\label{SubsecP2P}

Short video sharing and peer-to-peer streaming have been widely
cited as two key driving forces to Internet video distribution, yet
their development remains largely separated. The peer-to-peer
technology has been quite successful in supporting large-scale live
video streaming (e.g. TV programs like PPLive and
CoolStreaming) and even on-demand streaming
(e.g. GridMedia). Since each peer contributes its
bandwidth to serve others, a peer-to-peer overlay scales extremely
well with larger user bases. YouTube and similar sites still use the
traditional client-server architecture, restricting their
scalability.

Unfortunately, our YouTube measurement results suggest that using
peer-to-peer delivery for YouTube could be quite challenging. In
particular, the length of a YouTube video is quite short (many are
shorter than the typical connection time in a peer-to-peer overlay),
and a user often quickly loads another video when finishing a
previous one, so the overlay will suffer from an extremely high
churn rate. Moreover, there are a huge number of videos, so the
peer-to-peer overlays will appear very small.\footnote{A very recent
study on MSN Video \cite{huang07peervod} has suggested a
\emph{peer-assisted VoD}. We notice however that the statistics for
MSN Video are quite different from YouTube, and the technique has
yet to be substantially revised for YouTube.}

Our social network finding again could be exploited by considering a
group of related videos as a single large video, with each video in
the group being a portion of the large one. Therefore the overlay
would be much larger and more stable. Although a user may only watch
one video from the group, it can download the other portions of the
large video from the server when there is enough bandwidth and
space, and upload those downloaded portions to other clients who are
interested in them. This behavior can significantly reduce the
bandwidth consumption from the server and greatly increase the
scalability of the system.

Finally, note that another benefit of using a peer-to-peer model is
to avoid single-point of failures and enhance data availability.
While this is in general attractive, it is worth noting that timely
removing of videos that violate the terms of use (e.g.,
copyright-protected or illegal content, referred to by the
``Removed'' category in Section \ref{SubsecCate}) have constantly
been one of the most annoying issues for YouTube and similar sites.
Peer-to-peer delivery will clearly make the situation even worse,
which must be well-addressed before we shift such sites to the
peer-to-peer communication paradigm.


\section{Conclusion}\label{Conclusion}

This paper has presented a detailed investigation of the
characteristics of YouTube, the most popular Internet short video
sharing site to date. Through examining the massive amounts of data
collected in a 3-month period, we have demonstrated that, while
sharing certain similar features with traditional video
repositories, YouTube exhibits many unique characteristics,
especially in length distribution, access pattern, and growth trend.
These characteristics introduce novel challenges and opportunities
for optimizing the performance of short video sharing services.

We have also investigated the social network among YouTube videos,
which is probably its most unique and interesting aspect, and has
substantially contributed to the success of this new generation of
service. We have found that the networks of related videos, which
are chosen based on user-generated content, have both small-world
characteristics of a short characteristic path length linking any
two videos, and a large clustering coefficient indicating the
grouping of videos. We have suggested that these features can be
exploited to facilitate the design of novel caching or peer-to-peer
strategies for short video sharing.


\bibliographystyle{IEEEtran}
\bibliography{./IEEEabrv,./YouTube}

\end{document}